\documentclass[aip,apl,amsmath,amssymb,preprint,superscriptaddress]{revtex4-2}
\usepackage{graphicx}
\usepackage{amsmath}
\usepackage{dcolumn}
\usepackage{bm}
\usepackage{bbold}
\usepackage{tabularx}
\usepackage{xcolor}

\bibstyle{apsrev4-2}
\begin{document}
\title{Electro-Mechanical Tuning of High-Q Bulk Acoustic Phonon Modes at Cryogenic Temperatures}
\author{William Campbell}
\affiliation{ARC Centre of Excellence for Engineered Quantum Systems} \affiliation{ARC Centre of Excellence for Dark Matter Particle Physics, Department of Physics, University of Western Australia, 35 Stirling Highway, Crawley WA 6009, Australia}
\author{Serge Galliou}
\affiliation{FEMTO-ST Institute, Univ. Bourgogne Franche-Comt\'{e}, CNRS, ENSMM, 26 Rue de l’\'{E}pitaphe 25000 Besan\c{c}on, France}
\author{Michael E. Tobar}
\affiliation{ARC Centre of Excellence for Engineered Quantum Systems} \affiliation{ARC Centre of Excellence for Dark Matter Particle Physics, Department of Physics, University of Western Australia, 35 Stirling Highway, Crawley WA 6009, Australia}
\author{Maxim Goryachev}
\email{maxim.goryachev@uwa.edu.au}
\affiliation{ARC Centre of Excellence for Engineered Quantum Systems} \affiliation{ARC Centre of Excellence for Dark Matter Particle Physics, Department of Physics, University of Western Australia, 35 Stirling Highway, Crawley WA 6009, Australia}
\begin{abstract}
We investigate the electromechanical properties of quartz bulk acoustic wave resonators at extreme cryogenic temperatures. By applying a DC bias voltage, we demonstrate broad frequency tuning of high-Q phonon modes in a quartz bulk acoustic wave cavity at cryogenic temperatures of 4 K and 20 mK. More than 100 line-widths of tuning of the resonance peak without any degradation in loaded quality factor, which are as high as $1.73\times 10^9$, is seen for high order overtone modes. For all modes and temperatures the observed coefficient of frequency tuning is $\approx$ 3.5 mHz/V per overtone number $n$ corresponding to a maximum of 255.5 mHz/V for the $n = 73$ overtone mode. No degradation in the quality factor is observed for any value of applied biasing field.
\end{abstract}
\date{\today}
\maketitle
\begin{figure}
\centering
\includegraphics[width=0.5\textwidth]{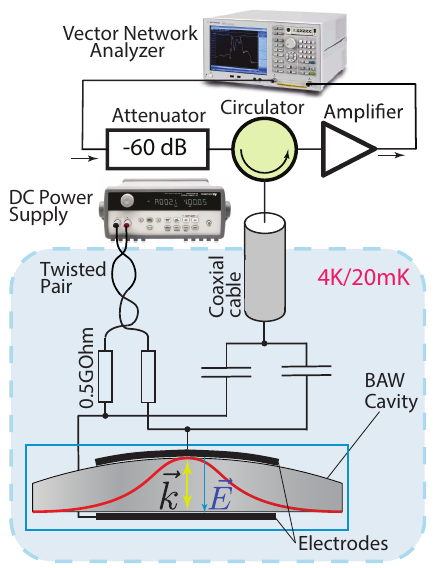}
\caption{Experimental setup for the excitation of acoustic modes in the quartz BAW resonator.}%
   \label{setupSIG}
\end{figure}
\section*{Introduction}
Low loss systems have found numerous applications as time-keeping devices as well as sensors, due to their long coherence times and high sensitivity. Although, in addition to high quality factors, many applications require a degree of frequency tunability. For example; frequency tuning is needed in clock applications to lock a low phase noise device to an external stable resonance or an atomic transition \cite{5116858,locke,Ivanov:2009pv,Ivanov1998,6174184}, or in detection systems where the frequency of a detectors target signal is unknown \cite{Campbell2021,Goryachev:2014ab}. However, it has been recognised that achieving both high tunability while maintaining high quality factors is extremely difficult both for electromagnetic and mechanical systems. In photonic applications this can be solved by changing geometry of the resonant system with mechanical actuators. In mechanical systems however, a geometrical tuning solution is almost impossible to realise without compromising losses. A lossless tuning mechanism for such mechanical resonators would be desirable as these devices may be more suitable than their optical counterparts for certain applications requiring lower frequency operation. In this work, we demonstrate highly stable electromechanical tuning of an acoustic quartz resonator, preserving the resonators extremely high quality factors.\\
Quartz bulk acoustic wave (BAW) resonators are primarily used in precision frequency and metrology applications as ultra-stable timing references, thanks to their extraordinarily high acoustic quality factors and frequency stability. For applications of these resonators that require locking to an external reference source, a traditional method of frequency tuning is to employ a tunable capacitor in the form of a varactor that can shift the resonant frequency by a few Hertz. The main drawbacks of this method are a low tuning range, additional noise due to the active component and an inability to work at cryogenic temperatures.\\
Another solution is to exploit the same piezoelectric coupling that is used to excite and read out the BAW modes, in order to change the geometry of the acoustic cavity. With efficient decoupling of the tuning and readout subsystems, this allows one to achieve broad frequency tunability without compromising system quality factors. Importantly, such a tuning mechanism is 
of great benefit when we consider the stable operation of quartz BAW resonators in a sub-Kelvin environment, where their performance significantly improves. Here they are known to exhibit an increase in acoustic quality factor $(Q)$ to as high as tens of billions, with a maximum reported\cite{ScRep} $Q\times f$ product of $7.8 \times 10^{16}$ Hz.\\
The design of a  mechanism to tune low loss cryogenic acoustic resonators, such as the quartz BAW, without degradation to $Q$ and without the need for spatially inefficient and thermally taxing components such as actuators, is highly motivated. Many applications such as quantum acoustics and qubit design \cite{Phys2020, Chu2017, Goryachev1, Woolley2016, Kharel2018}, development of cryogenic quartz oscillators with significant improvements to mid and long term frequency stability when compared to their room temperature counter-parts \cite{Goryachev2016}, as well as detection experiments searching for new physics such as dark matter or gravitational waves\cite{Campbell2021,Goryachev:2014ab,Goryachev:2018aa, Aggarwal2021, Goryachev2021, Manley2020}, benefit from increased sensitivity due to large $Q$s and effective resonator mode mass, as well as the low thermal noise of a cryogenic operating environment. Preserving quality factors while also introducing a degree of frequency tuning at cryogenic temperatures would be highly desirable to these applications. Additionally, the performance and characteristics of quartz BAW resonators operating at sub-Kelvin temperatures is relatively unexplored. This stark contrast to the well documented study of such resonators at room temperature motivates further investigation into system behaviours and device performance in the sub-Kelvin cryogenic regime.\\
\begin{figure}
\centering
\includegraphics[width = 0.5\textwidth]{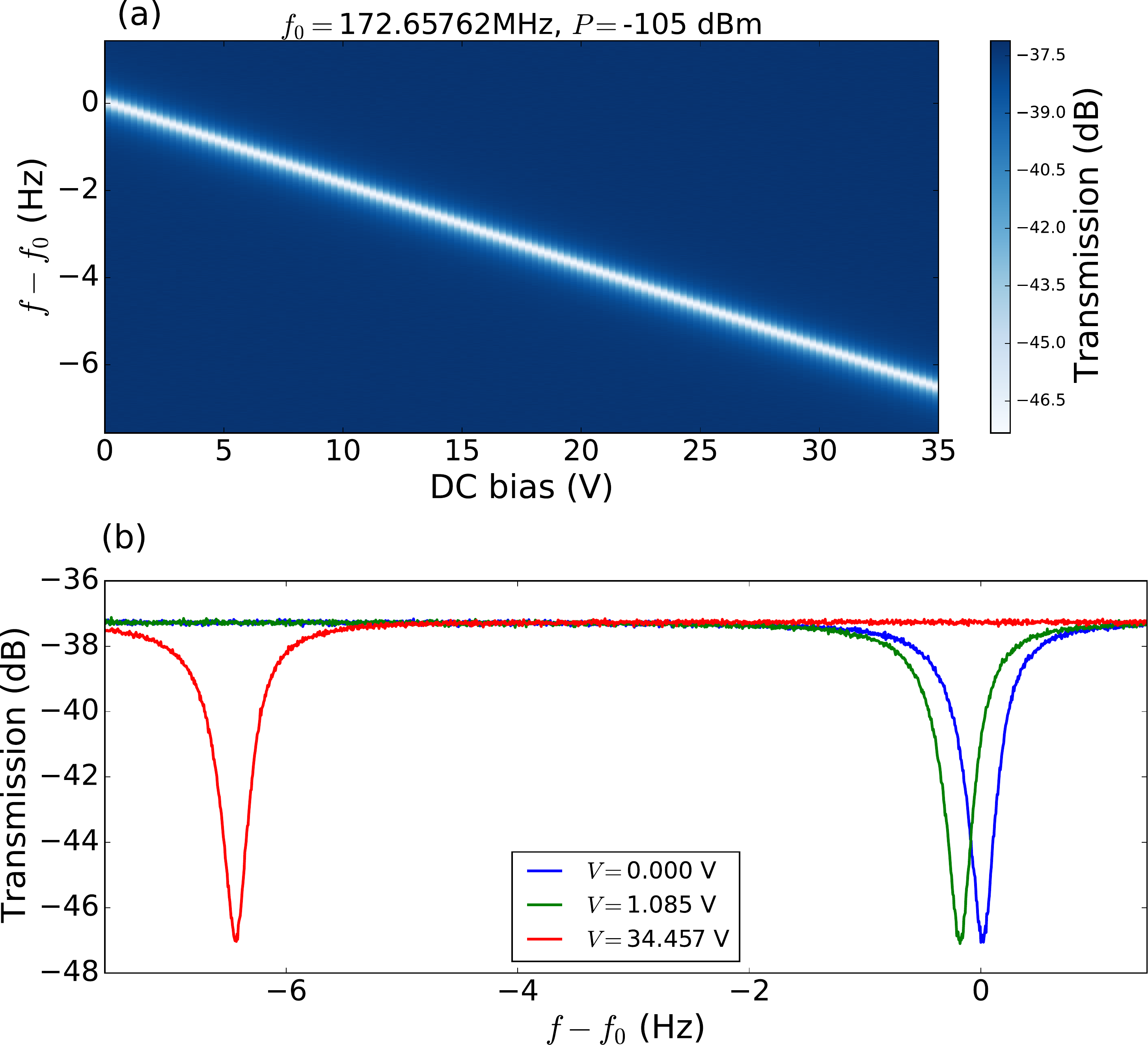}
\caption{Signal transmission through the experiment centred on the 55th overtone (173 MHz) mode of the quartz BAW at $T = 4$ K. (a) Density plot shows the system frequency response for varying applied bias field (b) Plotted traces show the transmission for different applied voltages.}
\label{fig:tuning229}
\end{figure}
\begin{figure}
\centering
\includegraphics[width = 0.5\textwidth]{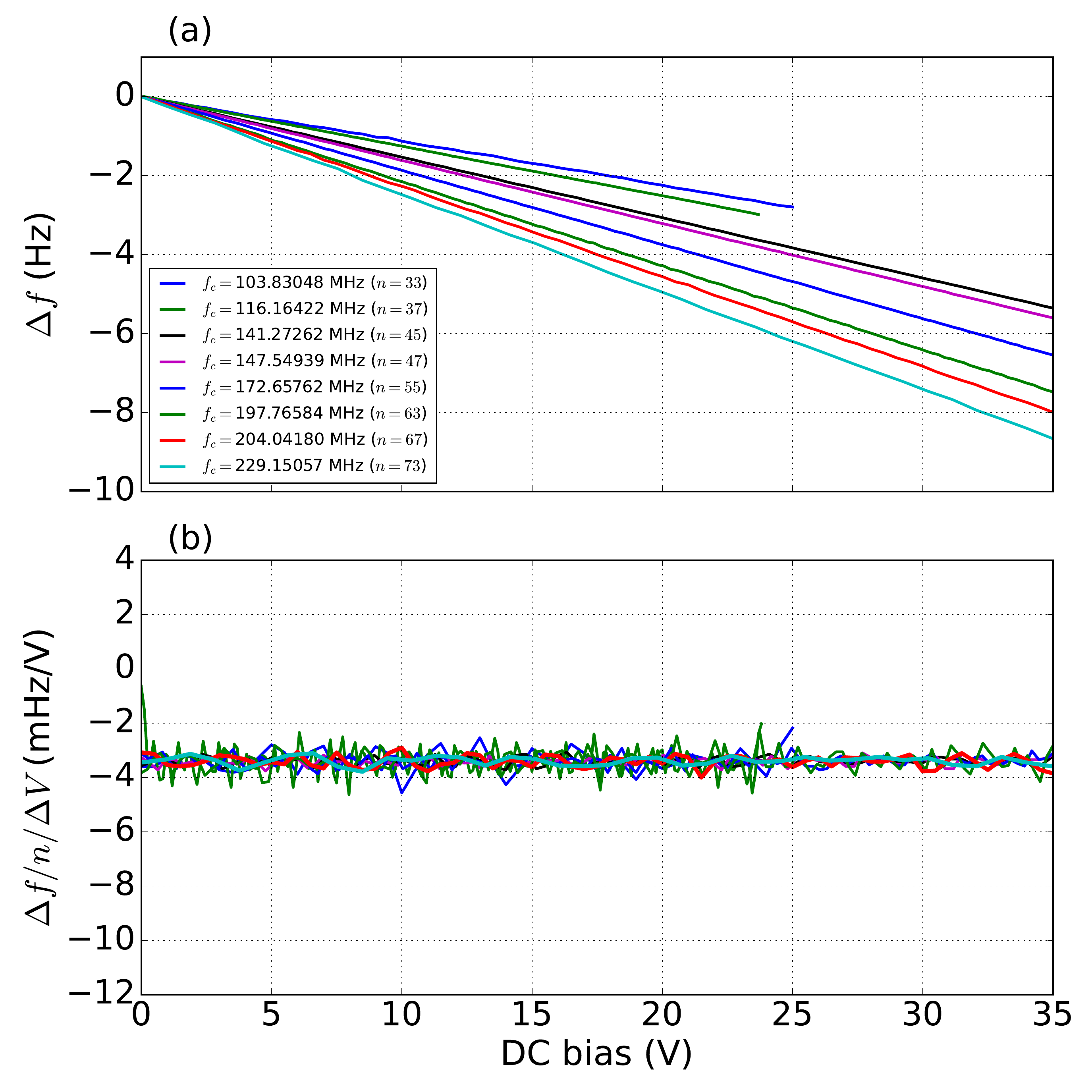}
\caption{Plotted traces show the measured frequency tuning of eight quartz BAW overtone modes at $T=$ 4 K for a varying applied bias field. (a) frequency of the resonators maximum response as a function of applied bias. (b) Frequency change normalised to mode number $n$ per change in applied bias for the same eight overtone modes.}
\label{fig:all modes}
\end{figure}
Previously, the loss mechanisms in quartz BAW resonators at both 4 K and milli-Kelvin has been studied \cite {Goryachev:2zn, apl1, apl2, Galliou2013, Galliou2014}, along with the relationship between mode frequency and temperature, where it was found that quality factors of BAW modes improve significantly at milli-Kelvin temperatures. In this work however, the relationship between external electric field and mode frequency is investigated, providing a different channel for studying these resonator systems. The broad frequency tuning capability of the quartz BAW resonator is demonstrated by making use of the piezoelectric effect to tune multiple phonon modes across a significant region of frequency space. The tuning mechanism is explored at both 4 K and 20 mK in order to confirm the technologies suitability as an ideal platform in the development of next generation low loss cryogenic resonator systems, primarily for fundamental physics tests.\\
\section{Experimental Methods}
A quartz plate 30 mm in diameter and 0.5 mm thick featuring an SC-cut \cite{Kusters:2014mn} crystal axis with a plano-convex \cite{Tiers2} surface was chosen for this work. The SC-cut axis of the quartz lattice allows for acoustic modes with better frequency stability and isolation from mechanical stresses, when compared to alternative axis cuts. The plano-convex surface concentrates the bulk phonon distribution to the center of the crystal, reducing anchoring losses and improving acoustic quality factors.\\
This crystal was situated in a copper vacuum enclosure that was placed into a dilution refrigerator where it was subject to a stabilised environment temperature $T$ of either 4 K or 20 mK. Acoustic modes of the quartz plate were excited by a signal supplied via microwave cables with capacitive DC blocks. These lines connect to two almost flat electrodes separated from the resonator crystal plate surface. The supplied microwave signal then couples to acoustic resonances in the crystal bulk via the piezoelectric effect. The set-up is displayed schematically in figure \ref{setupSIG}.\\
The crystal's resonant modes were observed by measuring the amplified reflection signal off the resonator with a vector network analyser (VNA) stabilized by a hydrogen maser. The use of a circulator was employed to isolate the reflected signal from the input source. In this configuration the reflected signal off the quartz can be amplified and measured as the $S_{21}$ transmission as seen by the VNA. The BAW excitation power is kept extremely low ($-105$ dBm) in order to avoid any inherent non-linear effects exhibited by the crystal at low temperatures.\\
\begin{figure}
\includegraphics[width = 0.5\textwidth]{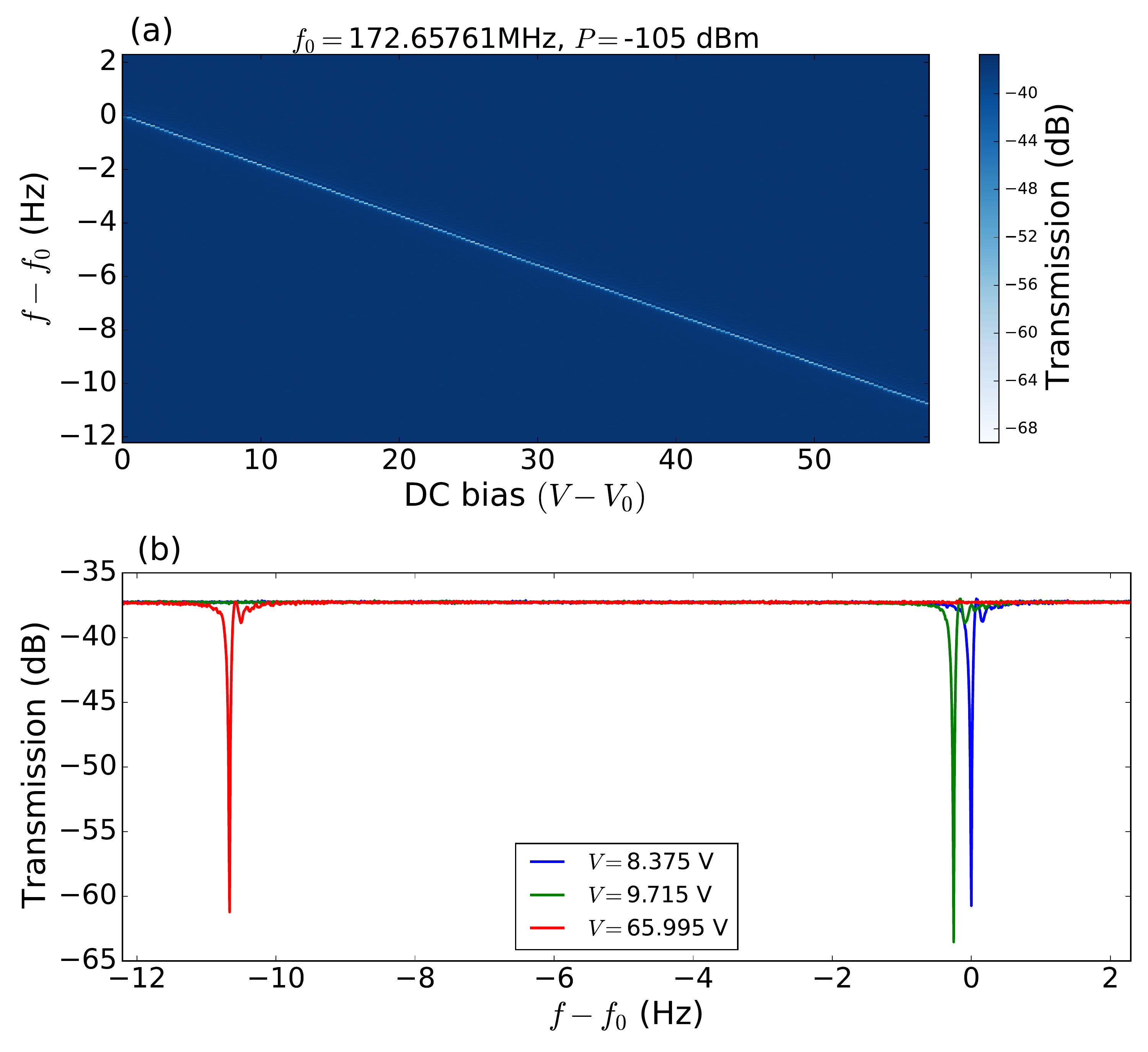}
\caption{Signal transmission through the experiment centred on the 55th overtone (173 MHz) mode of the quartz BAW at $T = 20$ mK. (a) Density plot shows the system frequency response for varying applied bias field (b) Plotted traces show the transmission for different applied voltages.}
\label{fig:tuning229mK}
\end{figure}
The piezoelectric coupling can be further exploited by applying an external DC electric bias field to the crystal in order to vary its bulk thickness, thus linearly shifting the resonant frequency of the acoustic modes. This can be thought of as the acoustic analogue of changing the length between the end mirrors in an optical Fabry-Perot cavity. To this end, an external power supply was connected to the electrodes so that an adjustable DC voltage would allow tuning of the resonator. Two cold $500~\text{M}\Omega$ resistances were used to isolate the power supply from the resonator in order to minimise loading of the resonator's intrinsic quality factor. Decoupling capacitors were used to separate the signal feeding lines and the ground from the external high voltage applied to the device.\\
\section{Results}
\begin{figure}
\includegraphics[width = 0.5\textwidth]{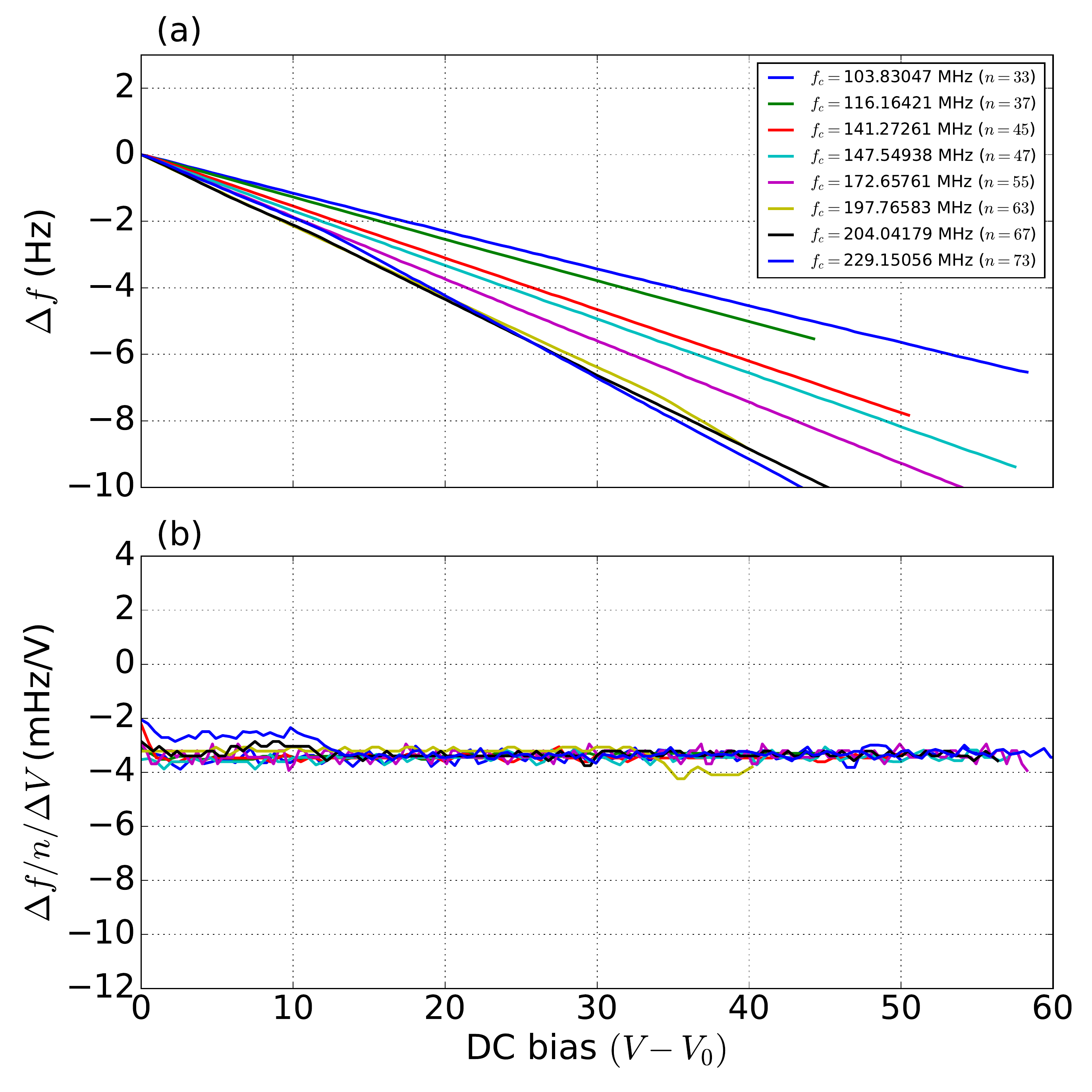}
\caption{\label{fig:all modes mK}Plotted traces show the measured frequency tuning of eight quartz BAW overtone modes at $T=$ 20 mK for a varying applied bias field. (a) frequency of the resonators maximum response as a function of applied bias. (b) Frequency change normalised to mode number $n$ per change in applied bias for the same eight overtone modes.}
\end{figure}

To explore the frequency tuning of the device the BAW cavity was held in the dilution refrigerator where it was subject to a stable external environment temperature of $T =$ 4 K. An external DC voltage supply was ramped in discrete stabilized steps from an initial value $V_0$ to a maximum of 67 V, applying a linearly varying static electric field to the quartz crystal. At each step the crystal response was measured with extremely narrow frequency resolution centred around a specific mode of the resonator. An IF bandwidth of 1 Hz was utilised, minimising the excitation signal sweep rate and allowing for ring-down effects to settle.\\
Eight various overtone modes where monitored with resonant frequencies ranging from 104 MHz to 229 MHz. In all cases consistent linear tuning of the corresponding frequency was observed, with no noticeable changes to the Lorentzian resonance shape or loaded quality factor, an example is given in figure \ref{fig:tuning229} where the initial bias field value $V_0$ is 0 V.\\
The change in frequency for all modes was found to strictly obey a $\Delta f\propto n$ dependence where $n$ is the overtone number. This relationship is clearly seen in figure \ref{fig:all modes}. This suggests that as is the case with this device at room temperature; applied electric field creates a strain across the crystal via the piezoelectric effect, changing  the thickness of the quartz plate and in turn affecting the resonant condition in the bulk as $f \propto \frac{n}{L}$ where $L$ is the length of the acoustic cavity. The relationship can be quantitatively described by a tuning coefficient of $\approx 3.5$ mHzV$^{-1}n^{-1}$.\\
In order to further these investigations; the above experiments where then repeated at the dilution refrigerators base temperature of 20 mK to determine if the same bulk thickness tuning mechanism can be sustained in a sub-Kelvin regime, where the quartz resonator is known to exhibit strong non-linear effects \cite{Goryachev2014jump}. In all cases across the same overtone modes, as seen in figure \ref{fig:all modes mK}, similar linear tuning proportional to the acoustic mode number was observed. The same $\approx 3.5$ mHzV$^{-1}n^{-1}$ tuning coefficient was extracted from these results where $V_0$ was set to $\approx$ 5 V in order to overcome an observed voltage suppression in superconducting DC feed lines. These results confirm consistent electromechanical coupling of the the crystal to electrically induced strain fields at extremely low cryogenic temperatures. While linear tuning may be a highly expected result for a room temperature device, it is of interest to see this behaviour adhered to in a milli-Kelvin environment, where the primary loss mechanisms in the crystal may differ and non-linear crystal properties become more dominant.\\
The highest magnitude of frequency tuning achieved was a $16.4$ Hz shift of the $n = 73$, 229 MHz mode at 20 mK when biased across a $\approx60$ V range. This corresponds to a controllable linear frequency shift of more than 30 line-widths of the resonant mode ($Q_{\textrm{loaded}}^{(n=73)} \approx~5.41 \times 10^8$). However, for higher Q modes the relative frequency shift is more substantial. For example the $n = 55$, 172.6 MHz mode at 20 mK, as depicted in figure \ref{fig:tuning229mK}, shifted 10.8 Hz under a similar applied voltage range, with $Q_{\textrm{loaded}}^{(n=55)} \approx 1.73 \times 10^9$. This corresponds to a frequency shift of 127 line-widths.\\
67 V was the maximum potential difference attainable in this work due to the current suitable equipment on hand, in principle larger bias fields are possible in order to demonstrate further tuning of the resonator. However, at some point extremely large strain fields will be at risk of damaging the quartz crystal plate. A summary of all the experimental findings is given in table \ref{tab:results}.\\
Comparing the performance of the presented device and tuning mechanism to other acoustic resonators, it is seen that smaller nano-mechanical flexural resonators are capable of greater frequency tuning whilst maintaining relatively good quality factors\cite{Zhang2015}. Such a typical device\cite{HBAR} exhibits 17 000 line-widths of tuning at mode center frequency $f_0=$ 6.5 MHz with quality factor $0.34\times 10^6$, thanks to an actuated applied stress tuning mechanism. However, the extremely low effective mode mass of these nano-scale resonators, as well as the inability to tune the device at cryogenic temperatures, makes them ill candidates for fundamental physics tests.\\
Thin-film and high overtone bulk acoustic wave resonators (FBAR/HBAR) are from the same family as the quartz BAW resonator and can be frequency tuned by changing the material dimensions or adding passive components much the same as quartz BAWs \cite{Liu2020}. These devices are typically micrometer-scale and designed for GHz frequency operation. Frequency tuning at the expense of quality factor can be gained by manufacturing FBAR resonators from ferroelectric materials, allowing the device to be tuned via an external electric field. A competitive HBAR resonator for example\cite{Berge2011} demonstrates $\approx$ 5 line-widths of tuning at $f_0 = 1.5$ GHz with a maximum maintained quality factor of just 130.\\
Centimetre-scale mechanical resonators, such as gravitational wave detector test masses \cite{ROWAN2005}, are ideal for fundamental physics tests due to their exceptionally high quality factors and cryogenic operating environment, however it is rare for such devices to display any degree of frequency tuning. The quartz BAW architecture thus provides competitive tuning ranges when compared to other resonators of the BAW family, with the added benefit of consistent linear tuning at cryogenic temperatures, as well as large effective mode mass and extraordinarily high quality factors. The combination of these traits make the quartz BAW device extremely suitable for precision tests of fundamental physics.
\begin{table*}[ht]
\centering
\begin{tabular}{lc|cc|cc}
 & & $T = 4$ K & & $T=20$ mK & \\
\hline
\hline
$n$ & $f_n$(MHz) & $\Delta f_n$ (Hz) & $V_\mathrm{max}-V_0$ (V) & $\Delta f_n$ (Hz) & $V_\mathrm{max}-V_0$ (V)  \\
\hline
\hline
33 & 103.83 & 2.79 & 25.0 & 6.53 & 58.29 \\
37 & 116.16 & 2.99 & 25.0 & 5.86 & 44.25 \\
45 & 141.27 & 5.35 & 35.0 & 8.02 & 56.5\\
47 & 147.55 & 5.60 & 35.0 & 9.46 & 57.5 \\
55 & 172.66 & 6.54 & 35.0 & 10.80 & 58.0\\
63 & 197.77 & 7.47 & 35.0 & 12.02 & 40.2\\
67 & 204.04 & 7.99 & 35.0 & 13.17 & 56.35\\
73 & 229.15 & 16.43 & 67.0 & 14.48 & 57.17\\
\hline
\hline
\end{tabular}
\caption{\label{tab:results}Summary of results for the tuning of all 8 overtone modes of the quartz BAW. The net frequency tuning of a mode resonance $f_n$ is denoted by $\Delta f_n$. The maximum applied bias field in each case is $V_\mathrm{max}-V_0$ where $V_0$ is an offset in order to overcome suppression effects associated with super conducting DC bias lines.}
\end{table*}
\section*{Conclusions}
To summarise, we have performed an investigation into the electro-mechanical properties of a quartz crystal BAW resonator in the sub-Kelvin regime, where we have observed a maximum of more than 100 line-widths of tuning of an acoustic resonance with no reduction to $Q$ or change in mode shape. We have further shown that this tuning mechanism is consistently linear and stable across a broad temperature range. The degree of tuning observed was completely limited by the equipment on hand, and the consistency of our results suggest that even further frequency tuning is indeed possible.\\
Many recent experiments that utilise these devices or similar resonators would significantly benefit from such frequency tuning. Precision experiments that use mechanical resonators to search for the small modulating signatures induced by ultra light dark matter \cite{Campbell2021, Manley2020, Arvanitaki2016, Carney2019} could utilise such frequency tuning in order to exclude a larger region of dark matter parameter space without sacrificing any sensitivity. This would bring mechanical resonator dark matter experiments in-line with the broad band frequency tuning capabilities of world leading cavity based searches such as ORGAN \cite{MCALLISTER201767} and ADMX \cite{Bartram2021}. 
High frequency gravitational wave search experiments \cite{Goryachev2021} could make use of this tuning mechanism in order to use advanced detection techniques such as cross correlation of the output of two identical quartz BAW detectors tuned to the same frequency. Such a system would be capable of probing a stochastic background of high frequency gravitational radiation \cite{Abbott2007}.\\  
This work presents an important step towards the stable operation of quartz crystal resonators at sub-Kelvin temperatures, where they have been known to display various instabilities and offer technical challenges \cite{Goryachev:2012jx, Goryachev:2013ly}. These investigations will aid in the development of highly tunable mechanical resonators for cryogenic operation, providing the tools required to perform high fidelity metrology, prepare mechanical systems for quantum ground state operation, and search for new physics with unparalleled sensitivity.\\
This research was supported by the Australian Research Council (ARC) Grant No. DP190100071, along with support from the ARC Centre of Excellence for Engineered Quantum Systems (EQUS, CE170100009) and the ARC Centre of Excellence for Dark Matter Particle Physics (CDM, CE200100008).\\
\section*{References}
\bibliography{BAW-DCtuning}

\begin{thebibliography}{10}

\bibitem{5116858}
Giorgio Santarelli, Graziano Governatori, Damien Chambon, Michel Lours, Peter
  Rosenbusch, Jocelyne Guena, Frederic Chapelet, Sebastien Bize, Michael~E.
  Tobar, Philippe Laurent, Thierry Potier, and Andre Clairon.
\newblock Switching atomic fountain clock microwave interrogation signal and
  high-resolution phase measurements.
\newblock {\em IEEE Transactions on Ultrasonics, Ferroelectrics, and Frequency
  Control}, 56(7):1319--1326, 2009.

\bibitem{locke}
C.~R. Locke, E.~N. Ivanov, J.~G. Hartnett, P.~L. Stanwix, and M.~E. Tobar.
\newblock Invited article: Design techniques and noise properties of
  ultrastable cryogenically cooled sapphire-dielectric resonator oscillators.
\newblock {\em Review of Scientific Instruments}, 79(5):051301, 2008.

\bibitem{Ivanov:2009pv}
E.~N. Ivanov and M.~E. Tobar.
\newblock Low phase-noise sapphire crystal microwave oscillators: current
  status.
\newblock {\em IEEE Transactions on Ultrasonics, Ferroelectrics, and Frequency
  Control}, 56(2):263--269, 2009.

\bibitem{Ivanov1998}
E.~N. {Ivanov}, M.~E. {Tobar}, and R.~A. {Woode}.
\newblock Microwave interferometry: application to precision measurements and
  noise reduction techniques.
\newblock {\em and Frequency Control IEEE Transactions on Ultrasonics,
  Ferroelectrics}, 45(6):1526--1536, November 1998.

\bibitem{6174184}
Jocelyne Guena, Michel Abgrall, Daniele Rovera, Philippe Laurent, Baptiste
  Chupin, Michel Lours, Giorgio Santarelli, Peter Rosenbusch, Michael~E. Tobar,
  Ruoxin Li, Kurt Gibble, Andre Clairon, and Sebastien Bize.
\newblock Progress in atomic fountains at lne-syrte.
\newblock {\em IEEE Transactions on Ultrasonics, Ferroelectrics, and Frequency
  Control}, 59(3):391--409, 2012.

\bibitem{Campbell2021}
William~M. Campbell, Ben~T. McAllister, Maxim Goryachev, Eugene~N. Ivanov, and
  Michael~E. Tobar.
\newblock Searching for scalar dark matter via coupling to fundamental
  constants with photonic, atomic, and mechanical oscillators.
\newblock {\em Phys, Rev. Lett.}, 126(7):071301, feb 2021.

\bibitem{Goryachev:2014ab}
Maxim Goryachev and Michael~E. Tobar.
\newblock Gravitational wave detection with high frequency phonon trapping
  acoustic cavities.
\newblock {\em Physical Review D}, 90(10):102005--, 11 2014.

\bibitem{ScRep}
S.~Galliou, M.~Goryachev, R.~Bourquin, Philippe Abbe, J.P. Aubry, and M.E.
  Tobar.
\newblock Extremely low loss phonon-trapping cryogenic acoustic cavities for
  future physical experiments.
\newblock {\em Nature: Scientific Reports}, 3(2132), 2013.

\bibitem{Phys2020}
F.~Souris, H.~Christiani, and J.~P. Davis.
\newblock {Tuning a 3D microwave cavity via superfluid helium at millikelvin
  temperatures}.
\newblock {\em Applied Physics Letters}, 111(17), 2017.

\bibitem{Chu2017}
Yiwen Chu, Prashanta Kharel, William~H Renninger, Luke~D Burkhart, Luigi
  Frunzio, Peter~T Rakich, and Robert~J Schoelkopf.
\newblock {Quantum acoustics with superconducting qubits}.
\newblock {\em Science}, 358(6360):199--202, 2017.

\bibitem{Goryachev1}
M.~Goryachev, D.~L. Creedon, E.~N. Ivanov, S.~Galliou, R.~Bourquin, and M.~E.
  Tobar.
\newblock Extremely low-loss acoustic phonons in a quartz bulk acoustic wave
  resonator at millikelvin temperature.
\newblock {\em Applied Physics Letters}, 100(24):243504, 2012.

\bibitem{Woolley2016}
M.~J. Woolley, M.~F. Emzir, G.~J. Milburn, M.~Jerger, M.~Goryachev, M.~E.
  Tobar, and A.~Fedorov.
\newblock Quartz-superconductor quantum electromechanical system.
\newblock {\em Phys. Rev. B}, 93:224518, Jun 2016.

\bibitem{Kharel2018}
Prashanta Kharel, Yiwen Chu, Michael Power, William~H. Renninger, Robert~J.
  Schoelkopf, and Peter~T. Rakich.
\newblock Ultra-high-q phononic resonators on-chip at cryogenic temperatures.
\newblock {\em APL Photonics}, 3(6):066101, 2018.

\bibitem{Goryachev2016}
Maxim Goryachev, Eugene~N. Ivanov, Michael~E. Tobar, and Serge Galliou.
\newblock Towards cryogenic quartz oscillators: Coupling of a bulk acoustic
  wave quartz resonator to a {SQUID}.
\newblock {IEEE}, may 2016.

\bibitem{Goryachev:2018aa}
M.~Goryachev, Z.~Kuang, E.~N. Ivanov, P.~Haslinger, H.~M{\"u}ller, and M.~E.
  Tobar.
\newblock Next generation of phonon tests of lorentz invariance using quartz
  baw resonators.
\newblock {\em IEEE Transactions on Ultrasonics, Ferroelectrics, and Frequency
  Control}, 65(6):991--1000, 2018.

\bibitem{Aggarwal2021}
Nancy Aggarwal, Odylio~D. Aguiar, Andreas Bauswein, Giancarlo Cella, Sebastian
  Clesse, Adrian~Michael Cruise, Valerie Domcke, Daniel~G. Figueroa, Andrew
  Geraci, Maxim Goryachev, Hartmut Grote, Mark Hindmarsh, Francesco Muia,
  Nikhil Mukund, David Ottaway, Marco Peloso, Fernando Quevedo, Angelo
  Ricciardone, Jessica Steinlechner, Sebastian Steinlechner, Sichun Sun,
  Michael~E. Tobar, Francisco Torrenti, Caner {\"{U}}nal, and Graham White.
\newblock {\em {Challenges and opportunities of gravitational-wave searches at
  MHz to GHz frequencies}}, volume~24.
\newblock 2021.

\bibitem{Goryachev2021}
Maxim Goryachev, William~M. Campbell, Ik~Siong Heng, Serge Galliou, Eugene~N.
  Ivanov, and Michael~E. Tobar.
\newblock Rare events detected with a bulk acoustic wave high frequency
  gravitational wave antenna.
\newblock {\em Phys, Rev. Lett.}, 127(7):071102, aug 2021.

\bibitem{Manley2020}
Jack Manley, Dalziel~J. Wilson, Russell Stump, Daniel Grin, and Swati Singh.
\newblock Searching for scalar dark matter with compact mechanical resonators.
\newblock {\em Phys. Rev. Lett.}, 124:151301, Apr 2020.

\bibitem{Goryachev:2zn}
M.~Goryachev, S.~Galliou, J.~Imbaud, R.~Bourquin, B.~Dulmet, and P.~Abbe.
\newblock Recent investigations on baw resonators at cryogenic temperatures.
\newblock {\em Frequency Control and the European Frequency and Time Forum
  (FCS), 2011 Joint Conference of the IEEE International}, pages 1--6, 2-5 May
  2011.

\bibitem{apl1}
Serge Galliou, Jo{\"e}l Imbaud, Maxim Goryachev, Roger Bourquin, and Philippe
  Abb{\'e}.
\newblock Losses in high quality quartz crystal resonators at cryogenic
  temperatures.
\newblock {\em Applied Physics Letters}, 98(9):--, 2011.

\bibitem{apl2}
Maxim Goryachev, Daniel~L. Creedon, Eugene~N. Ivanov, Serge Galliou, Roger
  Bourquin, and Michael~E. Tobar.
\newblock Extremely low-loss acoustic phonons in a quartz bulk acoustic wave
  resonator at millikelvin temperature.
\newblock {\em Applied Physics Letters}, 100(24):--, 2012.

\bibitem{Galliou2013}
Serge Galliou, Maxim Goryachev, Roger Bourquin, Philippe Abb{\'e}, Jean~Pierre
  Aubry, and Michael~E. Tobar.
\newblock Extremely low loss phonon-trapping cryogenic acoustic cavities for
  future physical experiments.
\newblock {\em Scientific Reports}, 3(1):2132, Jul 2013.

\bibitem{Galliou2014}
S.~Galliou, Ph. Abbe, R.~Bourquin, M.~Goryachev, M.~E. Tobar, and E.~N. Ivanov.
\newblock Properties related to q-factors and noise of quartz resonator-based
  systems at 4k.
\newblock {IEEE}, jun 2014.

\bibitem{Kusters:2014mn}
J.~A. Kusters.
\newblock The sc cut crystal - an overview.
\newblock {\em 1981 Ultrasonics Symposium}, pages 402--409, 14-16 Oct. 1981.

\bibitem{Tiers2}
H.F. Tiersten and D.S. Stevens.
\newblock An analysis of nonlinear resonance in contoured-quartz crystal
  resonators.
\newblock {\em J. Acoust. Soc. Am.}, 80(4):1122--1132, October 1986.

\bibitem{Goryachev2014jump}
Maxim Goryachev, Warrick~G. Farr, Serge Galliou, and Michael~E. Tobar.
\newblock Jump chaotic behaviour of ultra low loss bulk acoustic wave cavities.
\newblock {\em Applied Physics Letters}, 105(6):063501, 2014.

\bibitem{Zhang2015}
Wen-Ming Zhang, Kai-Ming Hu, Zhi-Ke Peng, and Guang Meng.
\newblock Tunable micro- and nanomechanical resonators.
\newblock {\em Sensors}, 15(10):26478--26566, 2015.

\bibitem{HBAR}
Johannes Rieger, Thomas Faust, Maximilian~J. Seitner, Jörg~P. Kotthaus, and
  Eva~M. Weig.
\newblock Frequency and q factor control of nanomechanical resonators.
\newblock {\em Applied Physics Letters}, 101(10):103110, 2012.

\bibitem{Liu2020}
Yan Liu, Yao Cai, Yi~Zhang, Alexander Tovstopyat, Sheng Liu, and Chengliang
  Sun.
\newblock Materials, design, and characteristics of bulk acoustic wave
  resonator: A review.
\newblock {\em Micromachines}, 11(7), 2020.

\bibitem{Berge2011}
John Berge and Spartak Gevorgian.
\newblock Tunable bulk acoustic wave resonators based on ba0.25sr0.75tio3 thin
  films and a hfo2/sio2 bragg reflector.
\newblock {\em IEEE Transactions on Ultrasonics, Ferroelectrics, and Frequency
  Control}, 58(12):2768--2771, 2011.

\bibitem{ROWAN2005}
S.~Rowan, G.~Cagnoli, P.~Sneddon, J.~Hough, R.~Route, E.K. Gustafson, M.M.
  Fejer, and V.~Mitrofanov.
\newblock Investigation of mechanical loss factors of some candidate materials
  for the test masses of gravitational wave detectors.
\newblock {\em Physics Letters A}, 265(1):5--11, 2000.

\bibitem{Arvanitaki2016}
Asimina Arvanitaki, Savas Dimopoulos, and Ken {Van Tilburg}.
\newblock {Sound of Dark Matter: Searching for Light Scalars with Resonant-Mass
  Detectors}.
\newblock {\em Physical Review Letters}, 116(3):1--6, 2016.

\bibitem{Carney2019}
Daniel Carney, Anson Hook, Zhen Liu, Jacob~M. Taylor, and Yue Zhao.
\newblock {Ultralight dark matter detection with mechanical quantum sensors}.
\newblock 2019.

\bibitem{MCALLISTER201767}
Ben~T. McAllister, Graeme Flower, Eugene~N. Ivanov, Maxim Goryachev, Jeremy
  Bourhill, and Michael~E. Tobar.
\newblock The organ experiment: An axion haloscope above 15 ghz.
\newblock {\em Physics of the Dark Universe}, 18:67--72, 2017.

\bibitem{Bartram2021}
C.~Bartram, T.~Braine, R.~Cervantes, N.~Crisosto, N.~Du, G.~Leum, L.~J.
  Rosenberg, G.~Rybka, J.~Yang, D.~Bowring, A.~S. Chou, R.~Khatiwada,
  A.~Sonnenschein, W.~Wester, G.~Carosi, N.~Woollett, L.~D. Duffy,
  M.~Goryachev, B.~McAllister, M.~E. Tobar, C.~Boutan, M.~Jones, B.~H. LaRoque,
  N.~S. Oblath, M.~S. Taubman, John Clarke, A.~Dove, A.~Eddins, S.~R. O'Kelley,
  S.~Nawaz, I.~Siddiqi, N.~Stevenson, A.~Agrawal, A.~V. Dixit, J.~R. Gleason,
  S.~Jois, P.~Sikivie, J.~A. Solomon, N.~S. Sullivan, D.~B. Tanner, E.~Lentz,
  E.~J. Daw, M.~G. Perry, J.~H. Buckley, P.~M. Harrington, E.~A. Henriksen, and
  K.~W. Murch.
\newblock Axion dark matter experiment: Run 1b analysis details.
\newblock {\em Phys. Rev. D}, 103:032002, Feb 2021.

\bibitem{Abbott2007}
B.~Abbott et~al.
\newblock First cross-correlation analysis of interferometric and resonant-bar
  gravitational-wave data for stochastic backgrounds.
\newblock {\em Phys. Rev. D}, 76(2):022001, jul 2007.

\bibitem{Goryachev:2012jx}
M.~Goryachev, S.~Galliou, P.~Abbe, P.~Bourgeois, S.~Grop, and B.~Dubois.
\newblock Quartz resonator instabilities under cryogenic conditions.
\newblock {\em Ultrasonics, Ferroelectrics and Frequency Control, IEEE
  Transactions on}, 59(1):21--29, January 2012.

\bibitem{Goryachev:2013ly}
Maxim Goryachev, Serge Galliou, Jo{\"e}l Imbaud, and Philippe Abb{\'e}.
\newblock Advances in development of quartz crystal oscillators at liquid
  helium temperatures.
\newblock {\em Cryogenics}, 57(0):104--112, 10 2013.

\end{thebibliography}
\bibliographystyle{unsrt}
\end{document}